\documentstyle[preprint,aps,epsfig]{revtex}
\begin{document}
\draft

\title{
       Radiation from a uniformly accelerating harmonic oscillator	}
\author{   
Hyeong-Chan Kim\thanks{me@taegeug.skku.ac.kr}   
}
\address{           
Dept. of Physics, Sungkyunkwan Univ., SUWON 440-746, KOREA
}
\author{
Jae Kwan Kim\thanks{jkkim@chep5.kaist.ac.kr} 
}
\address{           
Dept. of Physics, KAIST , Taejon 305-701, KOREA
}

\date{April 8, 1997}
\maketitle

\begin{abstract}

We consider a radiation from  a uniformly accelerating
harmonic oscillator  whose minimal coupling to the scalar field changes
suddenly.  The  exact time evolutions of the  quantum operators are given
in terms of a classical solution of a forced harmonic oscillator.
After the jumping of the coupling constant there occurs  a fast absorption 
of energy into the oscillator, and then  a slow emission follows.
Here the absorbed energy is independent of the acceleration and 
proportional to the log of  a  high momentum cutoff of the field.  
The  emitted energy  depends on the acceleration and also 
proportional to the log of the cutoff.
Especially, if the coupling is  comparable to the natural frequency of the 
detector ($e^2/(4m) \sim \omega_0$)  enormous energies are radiated
away from the oscillator. 

\end{abstract}
\begin{flushleft}
{ ~~~~~pacs number \hspace{.5cm}   04.60.+n,  03.70.+k  }
\end{flushleft}


\section{Introduction}

It is well known that the concept of a particle depends on 
the motion of an observer~\cite{birrell}. Especially, the Minkowski
vacuum is a canonical ensemble
with the temperature $a/2\pi$ from the point of view of a uniformly
accelerated observer with the acceleration $a$ (the thermalization
theorem)~\cite{unruh}. 
This observer dependence is most easily shown if one use
a particle detector model invented by Unruh~\cite{unruh}
and DeWitt~\cite{dewitt79}. It consists of an idealized point
particle  with internal energy levels labeled by  
$E$, coupled via a monopole interaction with a scalar field
$\phi$ (Unruh-DeWitt model). 
Following  these, many works emerged in the literature.
Letaw~\cite{Letaw} exhibited the stationary world lines, on which 
the detectors in a vacuum have a time-independent 
excitation spectra. 
Grove and Ottewill~\cite{Grove:Otte} studied the problem of 
a non-extended detector, and clarified the radiation effect 
arising both from the walls of the detector and from the
interaction with the field. Several authors~\cite{hinton,takagi}
discussed the anisotropic nature of the thermal radiation of the 
accelerated detector. A full review for this thermal character was 
given by  Takagi~\cite{takagi86}. The 
vacuum noise seen by a uniformly accelerated observer 
in flat space-times of arbitrary dimensions was investigated
and was shown to exhibit the phenomenon of the apparent inversion
of statistics in odd dimensions, which was discussed precisely by 
Unruh~\cite{unruh86} and Fukazawa~\cite{fukazawa}.
A few years ago, the excitation rate associated with a uniformly
accelerated finite-time detector interacting with the Minkowski
vacuum has been analyzed in an inertial frame by Svaiter and 
Svaiter~\cite{svaiter}.
They found a logarithmic ultraviolet divergences 
on the transition amplitude, which was  due to 
the instantaneous switching of the detector~\cite{higuchi}.
This UV divergence does not occur in lower dimensions. 
Grove argue that a  macroscopic constantly accelerating object
will emit negative energy radiation until
equilibrium with the Minkowski vacuum is achieved~\cite{grove}.

Several years ago a new particle detector model--a harmonic oscillator
coupled to a scalar field in $1+1$ dimensions--was introduced by Raine, 
Sciama, and Grove(RSG)~\cite{raine}. Several aspects of this model was 
discussed in connection with the `open quantum 
system'  \cite{unruhZurek,unruhWald,anglin}.  
Hinterleitner \cite{hin} and Massar, Parentani, and Brout~\cite{massar}
shown that there is a polarization cloud  which surrounds 
the detector at all times  and energy is  exchanged with it
locally.  Audretsch and M\"{u}ller~\cite{aud}
explored nonlocal pair correlations in accelerated detector. 
Recently stochastic aspects of this detector were discussed by
Raval, Hu, and Anglin~\cite{raval}. These  works mainly interested
on the asymptotic states with time independent coupling. 
In this paper we consider the intermediate region during the 
equilibrium  achieved between the detector and the field.
We show that  this  is not a simple energy absorption process
but there are two main stages after the two systems in contact with.
First stage is a fast absorption of energy of the oscillator from the field.
This occurs shortly after the change of the coupling in a  time 
which is much smaller than the inverse of the 
characteristic frequency of the oscillator. The total energy absorbed
during this period is independent of the acceleration and depends
on the log of a high momentum cutoff. 
Second stage is slow emission of radiation which exponentially decrease 
in a time scale of the coupling constant.  The total radiated energy during
this period depends on the acceleration of the oscillator. If the coupling
constant is small then the total radiation is smaller than the inertial one.
But if the coupling is comparable to the characteristic frequency,
enormous energies are radiated away from the oscillator.
In the case of a weakly coupled system, the absorbed energy 
during the first stage is larger than the emitted one during the second
stage.

In Sec. II-A, we describe the model in Minkowski space and give 
the general form of the solution for the field and the oscillator.
These evolutions of the operators are given by use of the inhomogeneous
solution $G(\omega,t)$ of a forced harmonic oscillator.  Similarly,
all physical quantities like the correlation function or the stress tensor 
can be expressed with this single function.
  In Sec. II-B, the model is generalized  
to incorporate the uniformly accelerating oscillators. Sec. III is devoted 
to present two solvable models.  
$G(\omega,t)$ is obtained in the asymptotic region.
We obtain the stress tensor in Sec. IV  when the detector 
is turned on suddenly.  
Sec. V is summary and discussions. There are two appendices 
where we describe the details of the calculation of the stress 
tensor.
\section{Models for the particle detector}

Let us consider a minimally coupled system of a massless real scalar 
field $\phi(t,x)$ in two dimensions and a detector of a harmonic oscillator 
$q(t)$ with mass $m$.
The action for this system is
\begin{eqnarray}\label{ac}
S &=& \int \mbox{d}x \mbox{d}t
    \frac{1}{2}\left\{ \left(\frac{\partial}{\partial t} 
	\phi(t,x) \right)^2 
	- \left( \frac{\partial}{\partial x} \phi(t,x)\right)^2
	\right\}  \\
  &+& \int d\tau  \left\{ \frac{1}{2} m 
		\left(\frac{d q(\tau)}{d \tau} \right)^2 
	-\frac{1}{2} m \omega_0^2 q^2(\tau) -e(\tau)q(\tau) 
	\frac{ d\phi}{d \tau} \left(t(\tau),x(\tau)\right) \right\} 
       . \nonumber
\end{eqnarray}
The oscillator follows the explicitly given path $(t(\tau),x(\tau))$  where
$\tau$ is the proper time of the oscillator along the path.
In this paper, we select two paths through which the oscillator
moves: the inertial and the uniformly accelerated.  

Varying Eq.~(\ref{ac}) with respect to $\phi(t,x)$ and $q(\tau)$ we 
get the Heisenberg equation of motion for  the field and the oscillator
\begin{eqnarray}
\Box \phi(t,x) &=& 
   \frac{de(\tau)q(\tau)}{d \tau} \delta (\rho),  \label{2}\\
m \left( \frac{d}{d \tau}\right)^2 q(\tau) 
   &+& m \omega_0^2 q(\tau)
   = - e(\tau)\frac{d \phi}{d \tau}(t(\tau),x(\tau)), \label{3}
\end{eqnarray}
where $\rho$ is an appropriate space coordinate which is orthogonal
to $\tau$ and the path of the oscillator can be represented as $\rho=0$.
Eq.~(\ref{2}) can be integrated to give
\begin{eqnarray} \label{phi:uv}
\phi(t,x) = \phi^0(t,x) + \frac{e(\tau_{ret})}{2} q(\tau_{ret}),
\end{eqnarray}
where $\tau_{ret}$ is the value of $\tau$ at the intersection of the
past lightcone of $(t,x)$ and the detector trajectory.
where 
we have used the explicit retarded propagator of a massless field
\begin{eqnarray}
G_{\mbox{ret}}(t,x;0,0) = \frac{1}{2} \theta(t+x) \theta(t-x).
\end{eqnarray}
Substituting the solution (\ref{phi:uv}) into Eq.~(\ref{3}),
one get 
\begin{eqnarray}\label{qeq}
m \ddot{q}(\tau) + \frac{1}{2} e^2(\tau) \dot{q}(\tau) + m \left( \omega_0^2
	+ \frac{\dot{e}^2(\tau)}{4 m}\right) q(\tau) 
	= -e(\tau) \dot{\phi}^0(t(\tau),x(\tau)).
\end{eqnarray}

The redefinitions 
\begin{eqnarray}
M(\tau) &=& m \exp\left(\int_{\tau_0}^{\tau}
	\mbox{d}\tau \frac{e^2(\tau)}{2m } \right), \label{M:t} \\
\omega^2(\tau) &=& \omega_0^2 + \frac{\dot{e}^2(\tau)}{4m}, \\
F(\tau) &=&  - \frac{e(\tau)}{m}  \frac{d\phi^0}{d\tau} (t(\tau),x(\tau)).
\end{eqnarray}
make Eq.~(\ref{qeq}) into the equation of motion of the forced harmonic
oscillator with the effective mass $M(t)$, and the 
frequency $\omega^2(t)$ 
\begin{eqnarray}\label{q'':F}
  \ddot{q}(\tau) + \frac{d\ln M(\tau)}{d \tau}\dot{q}(\tau) 
	+ \omega^2(\tau) q(\tau) = F(\tau).
\end{eqnarray}
Here $F(\tau)$ is the force density per unit effective mass.
We take the normalization of the effective mass as $M(\tau_0)= m$ at some
initial time $\tau_0$. As one see from  Eq. (\ref{q'':F}), we can 
arbitrarily  choose the normalization of the effective mass.
Note that we can rewrite this equation into quadratic form:
\begin{eqnarray} \label{quad}
\left[ \left(\frac{d}{d\tau}\right)^2 + \Omega^2(\tau)  \right] \sqrt{M(\tau)} q(\tau)
	= \sqrt{M(\tau)} F(\tau),
\end{eqnarray}
where 
\begin{eqnarray} \label{Omega}
\Omega^2(\tau) = \omega_0^2- \left( \frac{e^2(\tau)}{4m}\right)^2.
\end{eqnarray}
The behavior of a homogeneous solution of Eq.~(\ref{quad})
changes from oscillatory 
to exponential decay according to the value of $\Omega^2(t)$. We restrict
our  discussion into   $\Omega^2(t)$ greater than zero. 
If  $\epsilon(\tau) \ll \omega_0$ then $\Omega$  is natural positive
frequency mode of the oscillator.  The behavior of the  homogeneous 
solution,  in this case, is 
\begin{eqnarray}
f(t) = \frac{1}{\sqrt{M(\tau)}} exp\left[{\pm i \int^\tau \Omega(\tau') d\tau'}\right].
\end{eqnarray}

Let the initial Heisenberg operators for the oscillator 
to be $q(\tau_0)$ and $p(\tau_0)=m \dot{q}(\tau_0)$.  Then the 
exact quantum motion of $q(\tau)$ which is subjected to
the external force $F(\tau)$ in the Heisenberg picture is 
given by~\cite{kim} 
\begin{eqnarray}\label{q:0A}
q(\tau) &=& q_O(\tau)+ q_F(\tau) \nonumber \\
     &=& q(\tau_0)  \frac{\sqrt{g_-(\tau) g_+(\tau_0) }}{ \omega_I}  
        \cos \left[\Theta(\tau) - \chi(\tau_0)\right]
         + p(\tau_0) \frac{ \sqrt{g_-(\tau)g_-(\tau_0)}}{\omega_I} 
	\sin\Theta(\tau) \\
     &+& A_F(\tau) + A_F^{\dagger}(\tau). \nonumber
\end{eqnarray}
In this equations we use the following definitions:
\begin{eqnarray}
g_-(\tau) &=& f(\tau) f^*(\tau),   \label{g_-(t)}  \\
g_0(\tau) &=& - \frac{M(\tau)}{2} \dot{g}_-(\tau) ,  \nonumber \\
g_+(\tau) &=& M^2(\tau)\left|\dot{f}(\tau)\right|^2, \nonumber \\
	\label{phase}
\Theta(\tau) &=& \int^\tau_{\tau_0}\mbox{d}\tau
	 \frac{\omega_I}{M(\tau) g_{-}(\tau)}, 
\end{eqnarray}
where  $f(\tau)$ is a homogeneous solution of Eq.~(\ref{q'':F}) and 
$\omega_I = \sqrt{g_{+}(\tau) g_{-}(\tau) - g_{0}^2(\tau)}$ is 
invariant under the time evolution.  For 
the definition of $g_i(\tau)$ $(i= \pm,0)$ see \cite{kim} and 
references therein. If $\tau$ is Killing time, we can expand the 
free field solution into its positive solutions and negative solutions.
Let us classify its solution by $\omega$ and set the 
positive solution as $u_\omega$. Therefore
\begin{eqnarray}
\frac{\partial}{\partial \tau} u_\omega = -i |\omega|  u_\omega
\end{eqnarray}
and 
The free field solution in two dimension can be written as
\begin{eqnarray}
\phi^0(t,x) &=& \int_{-\infty}^{\infty} 
     \mbox{d}k [ a_k u_k(\tau, \rho) + a_k^{\dagger} 
	u_k^*(\tau,\rho) ] 
\end{eqnarray}
where $a_k$ and $a_k^\dagger$ is the corresponding  creation and
annihilation operators and $u_k$ is proportional to
$ e^{-i\omega \tau}$ .  With this choice, we can write the 
 annihilation part of the inhomogeneous solution 
$q_F(\tau)= A_F(\tau)+ A_F^{\dagger}(\tau)$  as
\begin{eqnarray}
A_F(\tau) = \int_0^{\infty} \mbox{d}\omega  \omega G(\omega,\tau)(
 	a_{\omega} + a_{-\omega}).
\end{eqnarray}
Where $G(\omega,\tau)$ is  the classical inhomogeneous solution of the forced
harmonic oscillator equation 
\begin{eqnarray}
\ddot{G}(\omega,\tau) + \frac{d \ln M(\tau)}{d\tau} \dot{G}(\omega,\tau) 
 	+ \omega^2(\tau)
	G(\omega,\tau) = -i \frac{e(\tau)}{m} u_\omega
\end{eqnarray}
with the initial condition 
\begin{eqnarray} \label{condition1}
G(\omega,0)=0  \hspace{1cm}   \dot{G}(\omega,0) = 0.
\end{eqnarray}

If we analyze $G(\omega,\tau)$, 
we can know all  time evolutions of the operators in principle.
The general solution for  $G(\omega,\tau)$ can be written as
\begin{eqnarray}\label{G:g}
G(\omega,\tau) = g(\omega, \tau) - g^*(-\omega,\tau),
\end{eqnarray}
where 
\begin{eqnarray} \label{g:tau}
g(\omega,\tau) = e^{i \Theta(\tau)}\frac{\sqrt{g_{-}(\tau)}}{2 m \omega_I} 
	\int_{\tau_0}^\tau \mbox{d}\tau'
		\sqrt{g_-(\tau')} M(\tau') e(\tau') e^{-i \Theta(\tau')}
		u_\omega (x(\tau'),t(\tau')).
\end{eqnarray}
One can show that Eq.~(\ref{q:0A}) satisfies (\ref{q'':F}) by direct
substitution.
The high momentum behavior of $G(\omega,t)$ is $O(1/\omega^{5/2})$
except some special case like the sudden jumping of the coupling
constant, which we consider   in Sec. IV.
In the case of large $\omega$ the integral of  (\ref{g:tau}) is approximately
given by the contributions around $\tau_0$, which makes the arguments
of the exponential of  $u_\omega(\tau)$ vanishes.
Therefore the first approximation of $g(\omega,\tau)$ is 
 of the form $\int d\tau' f(\tau_0) e^{\pm i \omega \tau}$.   But this term is 
canceled in  $G(\omega,\tau)= g(\omega,\tau) -g^*(-\omega,\tau)$, and
leaves only the $O(1/\omega^{5/2})$ and higher terms. 

\subsection{The inertial oscillator} \label{sec:II-1}

At first let us consider the simplest inertial path:  $x= 0$ and $t = \tau$.
Moreover the mode solution is $u_k = 1/\sqrt{4 \pi |k|} e^{-i (|k| t -k x)}$.

Let the initial state of the combined system to be 
\begin{eqnarray}\label{instate}
\left|i\right> =\left|n \right> \left|0\right>_M,
\end{eqnarray}
the $n$th excited state for the oscillator and the Minkowski 
vacuum state for the field.
The  correlation functions of $q(t)$  for state (\ref{instate}) is 
\begin{eqnarray}
\left<q_O(t) q_O(t')\right> &=&  (2n+1)
	   \frac{\sqrt{g_{-}(t)g_{-}(t')}}{2 \omega_I}
		\exp\left\{-i[ \Theta(t)- \Theta(t')] \right\},\\
\left<q_F(t) q_F(t')\right> &=& 2 \int 
     \mbox{d}\omega \omega^2 G(\omega,t) G^*(\omega, t'),\\
\left<q_O(t) q_F(t')\right> &=& 0. 
\end{eqnarray}
The correlation of the homogeneous part decrease because 
$M(t)$ increase monotonically.  Therefore for a large enough time
the correlation is governed by the inhomogeneous term.
If there is absent of $1/(\omega)^{3/2}$ term in $G(\omega,t)$ there
is no UV contribution to the correlation function. As we see in the 
previous section, this is normally true. Therefore  in the case of a
slowly varying coupling, the main contribution
to the correlation comes from the  frequency region around the 
resonance frequence $\Omega(t)$ (See Eq. (\ref{quad}) ).
The system is symmetric about $x=0$. Therefore, it is enough to
obtain the correlations of the field in the area $x,x'<0$.
In this area Eq.~(\ref{phi:uv}) becomes
\begin{eqnarray}
\phi(t,x) = \phi^0_R(u) + \phi^0_L(v) 
      + \frac{1}{2} e(v) [q_O(v)+q_F(v)].
\end{eqnarray}
Therefore the correlations of the field and the oscillator is for the 
state (\ref{instate}) are
\begin{eqnarray}
\left<\phi^0_R(u) q_F(v')\right> &=& \left<q_F(v') 
     \phi^0_R(u)\right>^* =
	\int \mbox{d}\omega \omega 
          G^*(\omega, v') u_{\omega}(u), \\
\left<\phi^0_L(v) q_F(v')\right> 
     &=& \left<q_F(v') \phi^0_L(v)\right>^* =
	\int \mbox{d}\omega \omega 
        G^*(\omega, v') u_{\omega}(v).
\end{eqnarray}
From these, one get the renormalized correlation function 
\begin{eqnarray}\label{correlation}
&&\left<\phi(t,x) \phi(t',x')\right> 
	  - \left<\phi^0(t,x)\phi^0(t',x')\right>    
	  = \frac{e(v)}{2} \left\{ \left<\phi^0_R(u) q_F(v')\right>
	    + \left<\phi^0_L(v) q_F(v')\right> 
	    \right\}  \nonumber \\
	&&~~~+ \frac{e(v')}{2}\left\{\left<q_F(v)\phi^0_R(u')\right> 
	    + \left<q_F(v)\phi^0_L(v')\right> \right\} 
          + \frac{e(v)e(v')}{4} \left\{\left<q_O(v) q_O(v')\right> 
	   + \left<q_F(v)q_F(v')\right> \right\} \nonumber \\
&&~=(2n+1) \frac{e(v)e(v')}{4} \frac{ 
	    \sqrt{g_{-}(v)g_{-}(v')}}{2 \omega_I}
            \exp\left\{-i[ \Theta(v)- \Theta(v')] \right\}  \\ 
&&~~+ \frac{e(v)e(v')}{2} \int \mbox{d}\omega 
	    \omega^2 G(\omega,v) G^*(\omega, v')  \nonumber \\
        &&~~+ \frac{e(v)}{2} \int \mbox{d}\omega\omega 
	     G^*(\omega,v') \left[ u_\omega(v) +u_\omega(u)\right] 
           + \frac{e(v')}{2} \int \mbox{d}\omega\omega 
	     G(\omega,v) \left[ u_\omega^*(v') +u_\omega^*(u') 
		\right]. \nonumber 
\end{eqnarray}

\subsection{ The uniformly accelerated oscillator}

 Now let us consider a uniformly accelerating trajectory
$x = \frac{1}{a} \cosh a\tau,  t= \frac{1}{a}\sinh a \tau $. 
Rindler space $(\tau,\rho)$ is given by
\begin{eqnarray}
x = \frac{1}{a} ~e^{a\rho} \cosh a\tau, 
~~ t =  \frac{1}{a} ~e^{a\rho} \sinh a \tau.
\end{eqnarray}
In this system the retarded time is 
\begin{eqnarray}
\tau_{ret} &=&\tau-\rho  ~~~\mbox{for} ~~ \rho > 0,\\
	   &=&\tau+\rho  ~~~\mbox{for} ~~ \rho < 0 .\nonumber 
\end{eqnarray}
At the right Rindler wedge, the free field $\phi^0(t,x)$ can be 
expanded with the normal modes of Rindler space-time as
\begin{eqnarray}
\phi^0(t,x) &=& \sum_{k=-\infty}^{\infty} [b_k \xi_k + H.C.]  \\
	&=& \sum_{\lambda =0}^{\infty}[b_\lambda \xi_\lambda(U) 
	    + b_{-\lambda} \xi_{\lambda}(V) + H.C.], \nonumber
\end{eqnarray}
where $U = \tau - \rho=-\mbox{ln}(-a u)/a $, 
$V = \tau + \rho= \mbox{ln}(av)/a$, and 
$\xi_\lambda = 1/\sqrt{4 \pi |\lambda|} e^{-i \lambda U}
  = 1/\sqrt{4 \pi |\lambda|} (-a u)^{i \lambda/a}$, and $b_\lambda, 
b_\lambda^\dagger$ is the creation and annihilation operator
in the Rindler spacetime. 
Therefore we can set  $u_\omega \rightarrow \xi_\lambda$ 
and $\omega \rightarrow \lambda$ in Sec. II.

Let us consider the initial state to be (\ref{instate}).
The expectation value of $q(\tau)$ for $|i>$  is zero.
The correlation functions  for $|i>$ are
\begin{eqnarray}
\left<q_O(\tau)\right.&&\left. q_O(\tau')\right>
	=  (2n+1) \frac{\sqrt{g_{-}(\tau)g_{-}(\tau')}}{2 \omega_I}
          \exp\left\{-i[ \Theta(\tau)- \Theta(\tau')] \right\},\\
\left<q_F(\tau)\right.&&\left. q_F(\tau')\right> 
	= 2 \int \mbox{d}\lambda \lambda^2 \left[ \left\{1
	    + N(\lambda/a)\right\}
	    G(\lambda, \tau) G^*( \lambda, \tau') \right. \\ 
	+&& \left. N(\lambda/a)
     G^*(\lambda, \tau)G(\lambda, \tau') \right], \nonumber \\
\left<\phi^0_R(U)\right.&&\left. q_F(\tau'_{ret})\right> 
	= \left<q_F(\tau'_{ret}) \phi^0_R(U) \right>^*  \\
	=&&
	\int \mbox{d}\lambda \lambda 
	\left[ \xi_{\lambda}(U) G^*(\lambda, \tau') \left\{
	1+N(\lambda/a) \right\} + \xi^*_{\lambda} (U)
	G(\lambda, \tau'_{ret}) N(\lambda/a) \right], \nonumber \\
\left<\phi^0_L(V)\right.&& \left.q_F(\tau'_{ret})\right> 
	= \left<q_F(\tau'_{ret}) \phi^0_L(V)\right>^* \\
	=&&
	\int \mbox{d}\lambda \lambda \left[\xi_\lambda(V) 
	G^*(\lambda, \tau_{ret}')
	 \left\{ 1+N(\lambda/a) \right\} + \xi_\lambda^*(V)
	 G(\lambda, \tau'_{ret}) N(\lambda/a) \right], \nonumber
\end{eqnarray}
where $N(\Omega) = 1/(e^{2\pi \Omega}-1)$.
We use the fact that the Minkowski vacuum is FDU thermal bath 
with temperature $a/(2\pi)$ to the accelerating observer.
The renormalized correlation function 
can be obtained with the same method of inertial case.
These correlation functions can be devided into two classes:
First is those of zero temperature and second is thermal contributions.
The high momentum behavior of the second is cut  off by the 
presence of the exponential in the denominator. Therefore 
it is evident that the first term dominate the correlation if 
there is UV divergences due to the existence of the $1/\omega$
term in $G(\lambda,\tau)$.  The sudden jump case of the coupling
is exactly that case.

\section{Exactly solvable models} \label{sec:solvable}

\subsection{Constant coupling}

The most easiest problem is, of course, the  case of constant coupling ($e(\tau)=e$).
In this case 
\begin{eqnarray}
M(\tau) &=&  \frac{e^2}{2} (\tau-\tau_0)   \\
f(\tau) &=& \frac{1}{\sqrt{m}} e^{\pm (i \Omega  +e^2/(2m))(\tau-\tau_0)} 
\end{eqnarray}
 where $\Omega ^2 = \omega_0^2 - \epsilon^2$.
$G(\omega, \tau)$ satisfies the following equation:
\begin{eqnarray}
\ddot{G}(\omega,\tau) + \frac{e^2}{2m} \dot{G}(\omega,\tau) + \omega_0^2
	G(\omega,\tau) = -i \frac{e}{m} u_\omega
\end{eqnarray}
The general solution to this equation is  given by
\begin{eqnarray} \label{sol1}
G(\omega,\tau) &=& a \exp(-\epsilon \tau)  
		\exp(i\sqrt{\omega_0^2-\epsilon^2} \tau+ \alpha)
		+  i e \chi(\omega) u_\omega (\tau,0) 
\end{eqnarray}
where
\begin{eqnarray}
\chi(\omega) = \frac{1}{m \left[\omega_0^2
		-\omega^2 -2i \epsilon \omega \right]}  
\end{eqnarray}
and  $\epsilon = e^2/4m$. The coefficients $a$ and $\alpha$ must be
chosen $G(\omega,\tau)$ to satisfy the initial condition (\ref{condition1}).
The first term exponentially decay therefore there remain only the second 
term asymptotically.  
This is exactly the same result with 
Massar, Parentani, and Brout \cite{massar}. 

\subsection{Turn on of the coupling} \label{sec:turnon}

The next example  is given by the coupling 
\begin{eqnarray}\label{cha}
 2\epsilon(\tau) =\frac{e^2(\tau)}{2 m} &=&  \frac{e_-^2}{4m}
	 \left(1- \tanh \frac{\tau}{d} \right)+
			 \frac{e_+^2}{4m}
			 \left(1+ \tanh \frac{\tau}{d}\right)  \\
    & =& \epsilon_{-}\left(1- \tanh \frac{\tau}{d}\right)+\epsilon_{+}
	\left(1+ \tanh \frac{\tau}{d} \right). \nonumber
\end{eqnarray}
The limit $d \rightarrow 0$ corresponds to the sudden jump
and $d \rightarrow \infty$ to the adiabatic one. Eq. (\ref{Omega}) 
\begin{eqnarray}
\Omega^2(\tau) = \omega_0^2 - \epsilon^2(\tau) =
     \frac{\omega_-^2}{2}\left(1- \tanh \frac{\tau}{d} \right) +
     \frac{\omega_+^2}{2}\left(1+ \tanh \frac{\tau}{d} \right) +
     \frac{(\epsilon_+- \epsilon_-)^2/4}{\cosh^2(\tau/d)}
\end{eqnarray}
has two limiting values $\omega_\pm^2 = 
\omega_0^2 -\epsilon_\pm^2$  at the positive and negative 
infinity.  These two values define a natural positive frequency
modes  of the oscillator in the past and the future asymptotic region.
The effective mass (\ref{M:t}) becomes
\begin{eqnarray}\label{mass}
M(\tau) = m \exp \int 2 \epsilon(\tau) \mbox{d}\tau 
     = m \left( \frac{\cosh \tau/d}{
	  \cosh \tau_0/d}\right)^{(\epsilon_+-\epsilon_-)d}
       \exp \left[ (\epsilon_+ +\epsilon_-)(\tau-\tau_0) \right].
\end{eqnarray}
From these we get the homogeneous solution for the classical 
equation of motion~(\ref{q'':F})  
\begin{eqnarray}\label{f:t}
f(\tau) = \frac{1}{\sqrt{M(\tau)}} e^{-i(\omega_+ + \omega_-)\tau/2} 
	\left(\cosh \frac{\tau}{d} \right)^{-i(\omega_+-\omega_-)d/2}
	~_2F_1(\alpha_-, \alpha_+;1-i \omega_- d; y),
\end{eqnarray}
where 
\begin{eqnarray}
y &=& \frac{1+ \tanh \tau/d}{2}, \\
\alpha_\pm &=& \frac{1 \pm 
	\sqrt{1+ (\epsilon_+-\epsilon_-)^2d^2}}{2} 
	+ i \frac{(\omega_+ - \omega_-)d}{2},
\end{eqnarray}
and $_2F_1$ is the hypergeometric function~\cite{morse}.
We choose Eq.~(\ref{f:t}) to be pure positive frequency mode at 
the past infinity. On the other hand, it becomes generally
mixture of the positive and negative modes at
the future:  
\begin{eqnarray}
\lim_{\tau \rightarrow -\infty} f(\tau) 
    &=& \frac{2^{i(\omega_+-\omega_-)d/2}}{
	\sqrt{M(\tau)}} e^{-i \omega_-\tau} ,  \label{f:-} \\
\lim_{\tau \rightarrow \infty} f(\tau) 
    &=& \frac{2^{i(\omega_+-\omega_-)d/2}}{
	\sqrt{M(\tau)}} \left( \alpha e^{-i \omega_+ \tau} 
	+ \beta e^{i \omega_+ \tau} \right).
	\label{f:+}
\end{eqnarray}
where 
\begin{eqnarray}
\alpha &=&\frac{\Gamma(1-i\omega_- d) 
      \Gamma(1-i\omega_-d -\alpha_- -\alpha_+)
	}{\Gamma(1-i\omega_-d -\alpha_-) 
      \Gamma(1-i\omega_-d -\alpha_+)} , \\
\beta &=&\frac{\Gamma(1- i\omega_-d)
      \Gamma(\alpha_- +\alpha_+ -1+i\omega_-d)
	}{\Gamma(\alpha_-)\Gamma(\alpha_+)}.  
\end{eqnarray}
The absolute squares of $\alpha$ and $\beta$ 
\begin{eqnarray}
|\alpha|^2 &=& \frac{1}{2}\frac{\omega_-}{\omega_+} \frac{
	\cosh\pi(\omega_+ + \omega_-)d 
         + \cos 2\pi x}{\sinh \pi \omega_- d
	\sinh \pi \omega_+ d}, \\
|\beta|^2 &=& \frac{1}{2}\frac{\omega_-}{\omega_+} \frac{ 
	\cosh\pi(\omega_+ - \omega_-)d 
   	  + \cos 2\pi x}{\sinh \pi \omega_- d
	\sinh \pi \omega_+ d}
\end{eqnarray}
satisfy a Bogolubov type relation 
\begin{eqnarray}
|\alpha|^2 -|\beta|^2 =\frac{\omega_-}{\omega_+},
\end{eqnarray}
where $x = \sqrt{1 + (\epsilon_+ -\epsilon_-)^2 d^2}/2$. 
The factor $\omega_-/\omega_+$ comes from the change 
of the  natural frequency  of the oscillator \cite{jyji2}.
At the present problem the initial homogeneous solution for the 
oscillator decays by $1/\sqrt{M(\tau)}$ factor so the  asymptotic 
form for large $\tau$ is given by $q_F$. Therefore 
the particle creation or other related topics must be discussed
with the inhomogeneous solution $G(\omega,\tau)$ with respect
to the positive frequency mode at the future asymptotic region.
Since our primary purpose is not the oscillator state but the radiation
from the oscillator, we do not discuss  it further.
In the adiabatic limit $\beta$ vanishes, on the other hand, 
in the sudden jump limit it becomes $(1-\omega_-/\omega_+)/2$.

From (\ref{f:t}) one get 
\begin{eqnarray}\label{g_-:on}
g_-(\tau) = f(\tau) f^*(\tau) 
  = \frac{|_2F_1(\alpha_-, \alpha_+;1-i \omega_-d;y)|^2}{M(\tau)},
\end{eqnarray}
and the invariant frequency $\omega_I= \omega_-$.
The integral of generalized frequency (\ref{phase}) is
\begin{eqnarray}\label{phase1}
\Theta(\tau) &=& \omega_I \int^\tau \mbox{d}\tau'
	   \frac{1}{|_2F_1(\alpha_-, \alpha_+;
		1-i \omega_-d;y)|^2} \nonumber \\
	&=&  \omega_I \int^\tau \mbox{d}\tau'\frac{1}{R^2(\tau)}
	- \omega_I \int^\tau \mbox{d}\tau'\frac{1}{R^2(\tau)}
	\left(1- \frac{R^2(\tau)}{|_2F_1(\alpha_-, 
	\alpha_+;1-i \omega_-d;y)|^2}
	\right)    \\
	&=& \theta(\tau) -\theta(\tau_0)
	   - \theta_f(\tau), \nonumber
\end{eqnarray}
where $R(\tau)$ and $\theta(\tau)$ are the absolute 
value and the real phase of
\begin{eqnarray}
R e^{-i \theta(\tau)}=\alpha e^{-i \omega_+ \tau}
		+\beta e^{+i \omega_+\tau},
\end{eqnarray}
and $\theta_f(\tau)$ approaches to some finite value
as $t \rightarrow \infty$.
Eqs.~(\ref{q:0A} and~(\ref{g_-:on})  shows $q_O(t)$ 
decreases exponentially for $\tau \gg d$.
Using these and Eq.~(\ref{G:g}) we get the asymptotic form 
\begin{eqnarray}\label{Ginfty:chi}
G(\omega,\tau) &=& i e_+  \chi(\omega) u_\omega(\tau,0) \\
 &-& \frac{1}{2 m \omega_I \sqrt{M(\tau)}} 
   \left[\left\{\alpha \chi_f(-\omega) - \beta^* 
		\chi_f(\omega)\right\} e^{-i \omega_+ \tau} +
	\left\{\beta \chi_f(-\omega) -\alpha^* \chi_f(\omega)
	\right\} e^{i \omega_+ \tau} \right]\nonumber
\end{eqnarray}
where 
\begin{eqnarray}
\chi(\omega) &=& \frac{1}{m [\omega_0^2 - \omega^2 -2i 
	\epsilon_+\omega]},
\end{eqnarray}
and $\chi_f(\omega) = \lim_{\tau\rightarrow \infty} \chi_f(\omega,\tau)$.
\begin{eqnarray} \label{chif}
\chi_f(\omega,\tau) &=& -e_+ 
     \int_{\tau_0}^\tau d\tau'\sqrt{M(\tau')} R(\tau') e^{-i\omega_+\tau'} 
     u_\omega(\tau' ,0)  \\
     & & \cdot \left[1- \frac{\sqrt{M(\tau')g_-(\tau')}}{R(\tau')}
		\frac{e(\tau')}{e_+} e^{i[\omega_+\tau'
		 - \theta(\tau')]} \right].  \nonumber
\end{eqnarray}
This result is similar with that of the constant coupling except 
$\chi_f(\omega)$  is determined by the integral.

\section{Stress Energy tensor in 
    the Sudden Jump Limit} \label{sec:jump}

Now let us study  the stress tensor of the scalar field 
in the presence of the oscillator.  We consider instant
switching process ($d \rightarrow 0$ limit of Sec.~\ref{sec:turnon}).
We solve this problem up to zeroth order on $d$ or 
$e^{-2|\tau|/d}$, where $\tau$ is the proper time
seen by the oscillator.  We calculate $G(\omega,\tau)$ 
without explicit choice of coordinates system because it is common 
both the inertial and  the uniformly accelerating oscillator. 
Let us set $\tau_0=-\infty$ and rescale the mass to be $M(0) =m$.
Similarly we also set $\Theta(0) =0$.

After carrying out the integral (\ref{chif}) explicitly
in the limit $|t| \gg d$  we get 
\begin{eqnarray}\label{G:-+}
G(\omega, \tau) &=& i e_- \chi_-(\omega) u_\omega(\tau,0),
		\hspace{5.9cm}	~~~~~ \mbox{for} ~~ \tau \ll -d, \\
      &=& G_\infty(\omega,\tau) + 
	\frac{i e_+}{2\sqrt{4 \pi \omega}}
	e^{-\epsilon_+ \tau} \left\{ \chi_+^*(-\omega) 
	e^{i\omega_+\tau}+\chi_+(\omega)e^{-i\omega_+\tau} 
	\right\}, \mbox{  for}~~ \tau \gg d, 
	\nonumber 
\end{eqnarray}
where  
\begin{eqnarray}\label{chip:chida}
\chi_+(\omega) = \chi_d(\omega)+\frac{e_-}{e_+}
	\chi_-(\omega),
\end{eqnarray}
and
\begin{eqnarray}
\chi_d(\omega) &=& \frac{1}{m \omega_+}
    \left[\frac{1}{\omega- \omega_+ +i \epsilon_+}
    - \frac{1- e_-/e_+  }{ 
    + \omega-\omega_++i(\epsilon_+ - 2/d)} \right.  \\
  &-&\left. \frac{e_-}{2e_+}
     \left(\frac{1}{\omega-\omega_-+i\epsilon_- }+ 
	\frac{1}{\omega +\omega_- +i \epsilon_-} 
      \right)\right],  \label{chi_d} \nonumber \\
\chi_-(\omega) &=& \chi_-^*(-\omega)=
 	\frac{1}{m}\frac{1}{\omega_0^2-\omega^2-
		2i \omega \epsilon_-}.
	\label{chi_a}
\end{eqnarray}
Here needs some remarks. All physical quantities like the coupling,
classical solution, and effective mass must  be continuous
at $\tau=0$.  This constraints demands the second term 
in $\chi_d$, which makes $G(\omega,\tau)$ to be quadratically
decrease for large $\omega$. 
$G_\infty(\omega,\tau) = i e_+ \chi(\omega,\tau) u_\omega(\tau,0) $ 
dominates the asymptotic form of $G(\omega,\tau)$  and the second term, 
which is the effect of the change of the coupling, decrease exponentially
on the time scale of $1/\epsilon_+$. At $\tau <0$  the inhomogeneous
solution $G(\omega,\tau)$ is that of the equilibrium. Therefore our system 
represent a system which is in equilibrium at $\tau<0$ become dynamic
due to the change of the coupling at $\tau=0$. The solution 
$G(\omega,\tau) $describe this  dynamic approaching process to
 equilibrium.
 
If we restrict the region of $\omega$ as $0 < \omega < \Gamma \ll 1/d$,
 we can ignore the $d$ dependent term in $\chi_d(\omega)$.  In this limit
$\chi_d(\omega)$ becomes $O(1/\omega)$ and gives  cutoff dependent UV
behaviors.
On the other hands  in the case of $\Gamma \gg 1/d$,  $\chi_d(\omega)$
is $O(1/\omega^2)$ which makes the theory UV finite. But
we cannot get a sensible theory because the  asymptotic form 
of the stress tensor crucially depends on $1/d$, which is unphysical.
Therefore we restrict the cutoff $\Gamma \ll 1/d$ and also restrict 
our attention to $|\tau| > 1/\Gamma$.

\subsection{The Stress tensor in the presence of a inertial 
	oscillator}  \label{sec:stressiner}

In Sec. (\ref{sec:II-1}) we obtain the renormalized 
correlation function (Eq. (27)) in the region $x , x'<0$.
In this region there is no $u, u'$ dependent terms. Therefore
$T_{uu}$ component of the stress tensor vanishes. Moreover,
$\left<T_{uv}\right> = tr T/4 = 0$ since we are dealing with 
a massless field in two dimension and there is no trace 
anomaly because the curvature is zero.

The stress tensor vanishes in the region $v<0$
since $G(\omega,t)$ has the same form with the asymptotic
case (\ref{Ginfty:chi}) and there is no radiation asymptotically.

In the region $t > 0$ we must calculate the stress tensor
explicitly.   We restrict our attention to $v > 1/\Gamma \gg d$
since we are interested in the radiation after turn on 
the coupling. 
After taking  differentiation of  the correlation function with respect
 to $v$ and $v'$ followed by the limit $v' \rightarrow v$ we get 
\begin{eqnarray}\label{Tvv:G}
T_{vv} &=& T_1 + \frac{e_+}{2} T_2 +
      \frac{e_+^2}{2} T_3,
\end{eqnarray}
where
\begin{eqnarray}
T_1 &=& (2n+1) \frac{e_+^2}{8 \omega_I} \lim_{v'\rightarrow v}
   \left(\partial_v \partial_v'\sqrt{g_-(v)g_-(v')} 
    \exp \left[-i\left\{
    \Theta(v) - \Theta(v')\right\}\right] \right)  \\
T_2 &=&	\int d \omega \omega \left[ \partial_v 
		G^*(\omega,v) \partial_v
	    u_\omega(v) + \partial_v G(\omega,v)\partial_v
	    u_\omega^*(v)  \right],  \\
T_3 &=& \int d \omega \omega^2 \partial_v G(\omega,v)
		\partial_v G^*(\omega,v).
\end{eqnarray}
Where we ignore terms related with $\dot{e}(v)$ which
is important for $t \sim d$. Since we consider only the region
$t > 1/\Gamma \gg d$, it is safe to ignore such terms. 

Now let us write down only the dominant terms of the stress tensor. 
(For details see appendix.)
For small $v$, it is dominant
the interference ($T_2$) between the oscillator and the field.
\begin{eqnarray}
T_{vv}= -\frac{e^2_+ \omega_0}{8 \pi m \omega_+} e^{-\epsilon_+ v} 
	\left(1- \frac{e_-}{e_+}
	\right) \frac{1}{v} \cos (\omega_+v + \theta) , \hspace{0.5cm}
		\mbox{for} \hspace{0.3cm}  
		\frac{1}{\omega_0} \gg v > \frac{1}{\Gamma}.
\end{eqnarray}
In this region the energy is absorbed  
into the oscillator from the  field with the amount 
\begin{eqnarray} \label{Eabs}
E_{absorbed} = \frac{e^2_+}{8 \pi m } \left(1- \frac{e_-}{e_+}\right)
	\ln\frac{\Gamma}{\omega_0} + \mbox{smaller terms}.
\end{eqnarray}
For  $v \gg 1/\Gamma$, energy is radiated away
from the oscillator and $T_3$ is dominant.
\begin{eqnarray} \label{Tvvasym}
T_{vv} &=& \frac{e_+^4}{8 \pi} \left(\frac{\omega_0}{m \omega_+} \right)^2 
		e^{-2\epsilon_+ v} \cos^2(\omega_+ v + \theta)
		\left(1-\frac{e_-}{e_+}\right)^2 \mbox{ln}\left(\Gamma/\omega_0\right)
	\hspace{0.5cm} \mbox{for} \hspace{.3cm}  v \gg \frac{1}{\Gamma}.
\end{eqnarray}
where $\tan \theta = \epsilon_+/\omega_+$.
The total radiated energy in this region is
\begin{eqnarray} \label{Erad}
E_{radiated} = \frac{e^2_+}{8 \pi m } \left(1- \frac{e_-}{e_+}\right)^2
	\ln\frac{\Gamma}{\omega_0}
\end{eqnarray}
Therefore we can conclude that in general the absorbed energy into
the oscillator  is greater than the radiated one. 

\subsection{The Stress tensor in the presence of a uniformly 
	accelerating oscillator}  \label{sec:stressacc}

With the same reason given at the previous subsection, 
$T^A_{uu}$ and $T^A_{uv}$  are zero. Similarly, 
the stress tensor vanishes for $ V < 0$.

In the region $V>0$, we also restrict  to $V > 1/\Gamma \gg 1/d$, 
and we ignored $\dot{e}(\tau)$ related terms.
In this region, $T^A_{vv}$ components can be written as follows:
\begin{eqnarray}
T^A_{vv} =  T^A_{1}+ \frac{e_+}{2} T^A_{2} + 
	\frac{e_+^2}{2} T^A_{3},
\end{eqnarray}
where
\begin{eqnarray}
T^A_{1} &=& (2n+1)\frac{e_+^2}{8 \omega_I} 
	\lim_{v' \rightarrow v} \left(
         e^{-a (V + V')} \partial_V \partial_V'
	 \sqrt{g_-(V)g_-(V')} 
    \exp \left[-i\left\{
    \Theta(V) - \Theta(V')\right\}\right] \right)  \\
T^A_{2} &=& \int d \lambda \lambda  \left\{1 + 2 N\left(
	\frac{\lambda}{a} \right) \right\} 
   \left[\partial_V \xi_{\lambda}(V) \partial_V G^*(\lambda,V)
        + \partial_V \xi_{\lambda}^*(V) \partial_V G(\lambda,v)
   \right] e^{-2a V},  \\
T^A_{3} &=& \int d \lambda \lambda^2 \left\{1 + 2 N\left(
	\frac{\lambda}{a} \right) \right\}
	\partial_V G(\lambda,V) \partial_V 
	G^*(\lambda, V) e^{-2 a V}.
\end{eqnarray}
For small $V$  the stress tensor
is dominated by $T_2^A$ which is given by
\begin{eqnarray}
T_{vv} ^A &=& \frac{e^2_+ }{8 \pi m \omega_+} 
		\left(1- \frac{e_-}{e_+}
	\right) \frac{a}{\ln av }    \\
      & &\cdot \left\{ \beta_+ (av)^{i \frac{
	\omega_+}{a}} + \beta_+^* ( av)^{-i \frac{\omega_+}{a}}
	\right\} (av)^{-2- \epsilon_+/a} 
	\hspace{.5cm} \mbox{for} \hspace{.2cm} \frac{1}{\omega_0}
	 \gg  V > \frac{1}{\Gamma}, \nonumber
\end{eqnarray}
This is exactly the same form with the inertial oscillator except the 
retardation factor $(a v)^{-2}$ due to the acceleration and the mere coordinate
change $v \rightarrow \ln av/a$.
The total absorbed energy is given by
\begin{eqnarray}
E^A_{absorbed} = \frac{e^2_+}{8 \pi m } \left(1- \frac{e_-}{e_+}\right)
	\ln\frac{\Gamma}{\omega_0} + \mbox{smaller terms}.
\end{eqnarray}
This is exactly the same with Eq. (\ref{Eabs}).  Physically, it is natural
because  there is no enough time the acceleration to act on the short time
interference.
For $V \gg 1/\Gamma$,  $T_3^A$ is dominant.
\begin{eqnarray}
T_{vv}^A =  \frac{e_+^2}{8 \pi m \omega_+} 
     \left( 1- \frac{e_-}{e_+}\right)^2
	\ln \frac{\Gamma}{ \omega_0} 
	\left\{ \beta_+ (av)^{i \frac{
	\omega_+}{a}} + \beta_+^* ( av)^{-i \frac{\omega_+}{a}}
	\right\}^2 (av)^{-2- 2\epsilon_+/a}
	  \hspace{.2cm} \mbox{for} \hspace{.2cm} V \gg \frac{1}{\Gamma},
\end{eqnarray}
This equation is quite similar to Eq.(\ref{Tvvasym}) except the 
retardation effect and the coordinate change
 ($v \rightarrow \ln{a v}/a $) due to acceleration.
This is because the main effect to the radiation comes
from the high momentum region.
The total energy radiated away from the oscillator is 
\begin{eqnarray}
E^A_{radiated} &=& \frac{e_+^2}{8 \pi m}
		\left(1-\frac{e_-}{e_+} \right)^2
	\ln \left(\frac{\Gamma}{\omega_0} \right) \tan \theta
	\left[\frac{1}{(x+ \sin \theta) \cos \theta} + 
	    \frac{\cos 2 \theta x - \tan \theta}{x^2 + 2 \sin \theta x + 1}
        \right] ,
\end{eqnarray}
where $x = a/(2 \omega_0)$.

\section{Summary and Discussion}

We  have discussed the influence of a harmonic oscillator on a
scalar quantum field in $1+ 1$ dimensions.  These are illustrated
by calculating the radiation of the scalar field from the oscillator. 
The first step to do this is to express the time evolutions with
the classical inhomogeneous  solution  $G(\omega,t)$ of a damping 
forced  harmonic oscillator. Then we applied this result to the 
sudden jumping limit of the coupling and obtain the change of the
stress tensor in the presence of the oscillator.  

There are two main effects on the radiation.
The first is  due to sudden change of the coupling which is  described
by the correlation between the oscillator and the field. This effect
rapidly die out but  the oscillator absorbs large energy from the 
field through this correlation. Moreover this absorption is independent of 
the acceleration.
Subsequently, slow radiation from the oscillator take place.
In case of an inertial oscillator this radiation is smaller than the absorbed
energy through the first stage.  The behavior of the total radiated 
energy become nontrivial if the detector is accelerated. 
In  the case of a small coupling constant ( $\epsilon_+ \ll \omega_0$),
 the radiated energy is maximized by $a=0$. But there is 
a peak  of the radiated energy at a non-zero acceleration 
if $\theta $ greater than some value $\theta_0 \sim 1.07702$ or
\begin{eqnarray}
  \left(\frac{\epsilon_+}{\omega_0}\right)^2  > 
	\frac{- \omega_+/\omega_0 + \sqrt{(\omega_+/\omega_0)^2 -
		8 \omega_+/\omega_0 +8}}{4(1-\omega_+/\omega_0)}
\end{eqnarray}
In this case the radiated energy can greater than the absorbed one
during the first stage. Especially, if $\epsilon \rightarrow \omega_0$
then the radiated energy becomes extremely large for non-zero
acceleration. 

If the acceleration is  large enough, the radiation decrease 
according to the inverse of the acceleration. The following two 
points can help to understand this phenomena. First, the radiation 
is not due to the acceleration but due to the change of the coupling.
Second,  as acceleration grows the unit proper time of an accelerating
oscillator correspond to a larger coordinate time to the Minkowski 
observer. Therefore the coordinate time which takes to vary the coupling
becomes larger for the larger acceleration.
\begin{figure}[bht]
\vspace{0.5cm}
\centerline{
\epsfig{figure=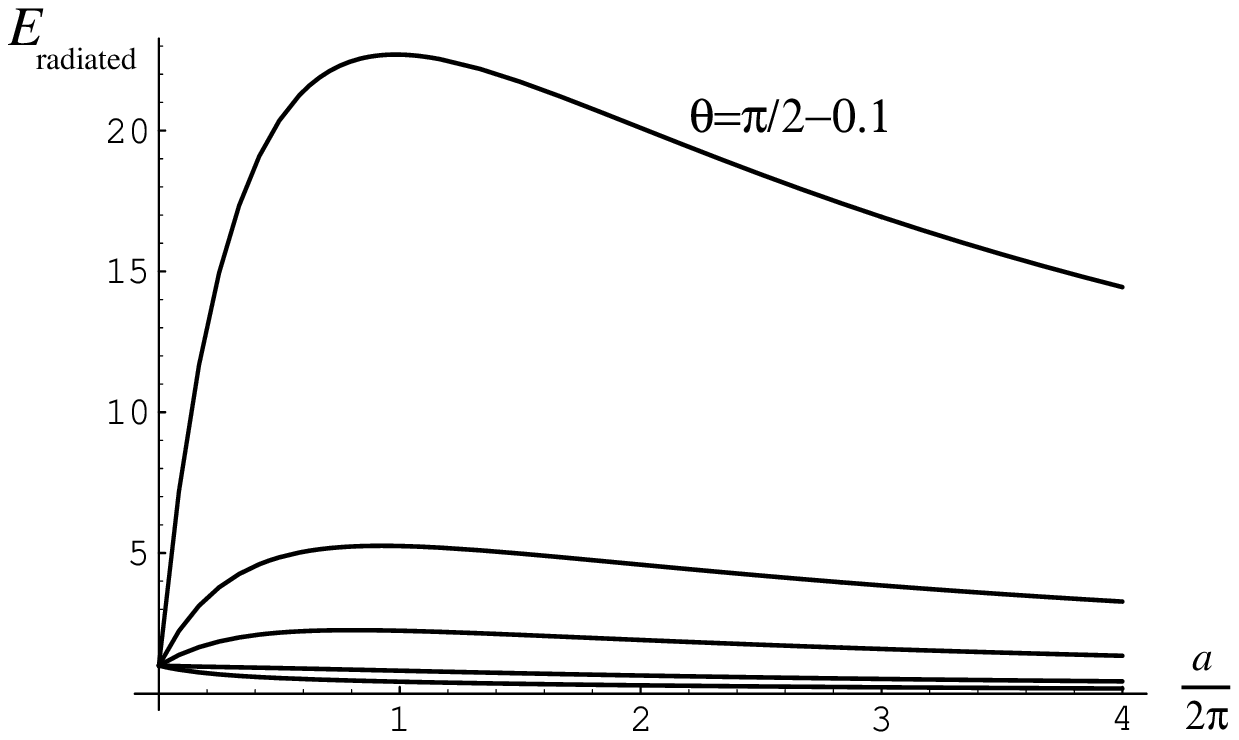,height=8cm}}
{\footnotesize Fig. 1.
Total radiated energy \\
Total radiated energy is plotted according to the acceleration.
The acceleration of  each time is given by $ a/(2\omega_0)= 
\{\pi/6, \pi/3, \pi/2-0.3, \pi/2-0.2, \pi/2-0.1\}$ from the below.
The unit for energy is
$\frac{e_+^2}{8 \pi m}	\left(1-\frac{e_-}{e_+} \right)^2
	\ln \left(\frac{\Gamma}{\omega_0} \right)$.
}
\end{figure}

~\\~\\~
\section{acknowledgements}
This work was supported by the
Korea Science and Engineering Foundation (KOSEF).
One of the authors thanks to   
Min-Ho Lee for his helpful discussions.

\section*{Appendix}
\subsection{The Stress Energy Tensor of the field 
	in the sudden jump of the coupling -- Inertial Case}

In this appendix we obtain the stress tensor for the model of 
Sec.~\ref{sec:turnon} in the sudden jump limit. It is easy 
to know that the stress tensor simply vanishes for $v<0$, 
from~(\ref{G:-+}). Therefore we calculate it only for 
$v \geq 0$ in the left hand side of the oscillator.  
The stress tensor~(\ref{Tvv:G}) is composed of three terms.

The first term can be evaluated easily to become
\begin{eqnarray} \label{T1}
T_1 &=&\frac{\epsilon_+}{4 \omega_{-}\omega_{+}^2} 
	e^{-2 \epsilon_+ v}
	\left[(\omega_+^2+ \omega_-^2)\omega_0^2
           +(\epsilon_+^2-\omega_+^2
	)(\omega_+^2-\omega_-^2) \cos 2\omega_+ v \right.  \\
     &&+ \left. 2\epsilon_+ \omega_+ (
	\omega_+^2- \omega_-^2) \sin2 \omega_+ 
		v \right]. \nonumber
\end{eqnarray}

$T_2$ is sum of two terms which are mutually complex conjugate.
One of these is 
\begin{eqnarray} \label{Gu}
\int d \omega \omega 
		\partial_v G(\omega,v)
	\partial_{v} u_\omega^*(v) = 
\int d \omega \omega 
	\partial_v G_\infty(\omega,v)
	\partial_{v} u_\omega^*(v)
	 +\frac{i e_+}{8 \pi}e^{-\epsilon_+ v} T_{2a},
\end{eqnarray}
where
\begin{eqnarray} \label{T2a}
T_{2a} = -\beta_+ e^{i\omega_+v} \int d \omega \omega
	\chi_+^*(-\omega) e^{i\omega v} 
	+ \beta_+^* e^{-i \omega_+v}
	\int d\omega \omega \chi_+(\omega) 
		e^{i \omega v} ,
\end{eqnarray}
and we define the constant
\begin{eqnarray}
\beta_\pm = \omega_\pm + i \epsilon_\pm.
\end{eqnarray}
Therefore
\begin{eqnarray}
T_{2a}-T_{2a}^* &=& -\frac{1}{m \omega_+} \beta_+ e^{i \omega_+ v}
    \left[ \left(1-\frac{e_-}{e_+}   \right)  \frac{2i}{v} + 2 \beta_+ e^{-i\beta_+ v}
    Ei(i \beta_+ v)   \right.  \\
    &-& \left. \frac{e_-}{e_+}\left\{ \left(1+ \frac{\omega_+}{\omega_-} \right)
	\beta_- e^{-i \beta_- v} Ei(i \beta_- v) - \left(1+ \frac{\omega_+}{\omega_-} \right)
	\beta_-^* e^{i \beta_-^*v} Ei(-i \beta_-^*v) 
        \right\}  \right]   \nonumber \\
    &-& C.C.  \nonumber
\end{eqnarray}
Rather than use this complex form, lets us extract only its limiting form
for small and large $v$ .
In the region $  d \ll 1/\Gamma  < v  \ll 1/\omega_0 $ 
\begin{eqnarray}
T_2 &=& \frac{e_+}{4 \pi m \omega_+} e^{-\epsilon_+ v}\left(1- \frac{e_-}{e_+}
	\right) \frac{ \omega_0}{v}  \cos (\omega_+v+ \theta)
\end{eqnarray}
and for large $v \gg  1/\omega_0$,  $T_2 = O(e^{-2 \epsilon_+ v})$.
Where $ \tan \theta = \epsilon_+/\omega_+$.

Finally, let us evaluate $T_3$.
If we define the following integrals
\begin{eqnarray}
I_1(v) &=& \int_0^\Gamma d \omega 
	 \omega^2 \chi(\omega) \chi_d(-\omega) e^{-i \omega v}, \\
I_2(v) &=& \int_0^\Gamma d \omega \omega^2 \chi(\omega)
	    \chi^*_d (\omega) e^{-i \omega v}, \\
J(v) &=&  \int_0^\Gamma d \omega \omega^2 \chi(\omega)
	  \chi_-(-\omega) e^{-i \omega v} ,
\end{eqnarray}
then  $T_3$ becomes
\begin{eqnarray}
T_3 = && \lim_{v' \rightarrow v} 
      \int d \omega \omega^2 \partial_v G_\infty(\omega,v) 
       \partial_{v'} G_\infty^*(\omega,v')    \\
     &+& \frac{e_+^2}{8 \pi} \left( 
       e^{-i\beta_+^* v} \beta_+^* T_{3a} + e^{i \beta_+ v}
	\beta_+ T_{3a}^* \right)+
	\frac{e_+^2}{16 \pi} e^{-2\epsilon_+ v} T_{3b}.
	\label{T3:T3ab} \nonumber
\end{eqnarray}
where
\begin{eqnarray}
T_{3a} &=& -I_1(v) + I_2^*(v)+
	  \frac{e_-}{e_+}[-J(v)+ 
	  J^*(v)]  
                       \label{T3a:int}  \\
T_{3b} &=& \omega_0^2 \int _0^\Gamma d \omega \omega
	  \left( |\chi_+(\omega)|^2+|\chi_+(-\omega)|^2
			\right)  \label{T3b:int} \\
    &&-\beta_+^2 e^{2i\omega_+ v} 
       \int _0 ^\Gamma d \omega \omega\chi_+^*(-\omega)\chi_+^*(
      \omega)-\beta_+^{*2} e^{-2i\omega_+ v}
	\int_0^\Gamma d \omega \omega 
	\chi_+(-\omega)\chi_+(\omega). 
	 \nonumber 
\end{eqnarray}
where we have introduced explicit high momentum cut-off  $\Gamma$ 
to regularize the UV behaviors.
The first term of $T_3$
is canceled by the $\lim_{v\rightarrow v'}\int d \omega \omega 
(\partial_vG_\infty \partial_{v'}u^*
+\partial_v u^* \partial_{v'}G_\infty)$
term of $T_2$ (The detail of the calculation can be consulted
in Ref. \cite{massar}.)  As one can see in $T_{3b} $ major contribution 
to the stress tensor comes from the ultra-violet region. As one can
easily see $T_{3b}$ don't have UV contribution.  Therefore
the major contribution comes from $T_{3b}$.
Let us examine $T_{3b}$ in detail.
$\chi_-$ is of order $O(1/\omega^2)$ for large $\omega$, 
therefore only the first term of the integral
\begin{eqnarray}\label{intchip}
\int d&& \omega \omega |\chi_+(\omega)|^2 \\
   &&= \int d \omega \omega |\chi_d|^2 
   +\frac{e_-}{e_+} \int d \omega \omega 
	\left( \chi_d \chi_-^* 
   + \chi_-\chi_d^* \right)
   +\frac{e_-^2}{e_+^2}
	 \int d \omega \omega |\chi_-|^2  \nonumber
\end{eqnarray}
can have important  ultra-violet  contribution.
If one try to extract only the high momentum part  it is ease to show that
\begin{eqnarray}
\int d \omega \omega |\chi_d|^2  &\cong & \frac{1}{m^2 \omega_+^2}
	\left(1-\frac{e_-}{e_+} \right)^2 \int^\Gamma d\omega \frac{1}{\omega} 
			\\
    &=& \frac{1}{m^2 \omega_+^2} \left(1-\frac{e_-}{e_+} \right)^2
		\mbox{ln}\left(\Gamma/\omega_0\right).  \nonumber
\end{eqnarray}
Therefore
\begin{eqnarray}
T_3 = \frac{e_+^2}{4 \pi} \left(\frac{\omega_0}{m \omega_+} \right)^2 
		e^{-2\epsilon_+ v} \cos^2(\omega_+ v + \theta)
		\left(1-\frac{e_-}{e_+}\right)^2 \mbox{ln}\left(\Gamma/\omega_0\right).
\end{eqnarray}
where $\tan \theta = \epsilon_+/\omega_+$.

\subsection{The stress tensor of the field in the sudden
	jump limit of the coupling -- Accelerating Case}

The stress tensor for $V<0$ vanishes. 
In the region $V>0$, the term $T^A_{1}$ is 
\begin{eqnarray}
T^A_1 (v) &=& \frac{\epsilon_+}{4 \omega_- \omega_+^2} 
	\left[ (\omega_+^2+ \omega_-^2)\omega_0^2  \right.\\
    &-& \left. \frac{1}{2}
	(\omega_+^2-\omega_-^2)
	\left(\beta_+^{*2} (av)^{-2i\omega_+/a}
	+ \beta_+^2 (av)^{2i \omega_+/a} \right) \right]
	(av)^{-2(1+\epsilon_+/a)}. \nonumber
\end{eqnarray}

The integral for $T^A_{2,3}$ are of the form
$\int d \lambda \lambda (1 + 2 N(\lambda/a) )f(\lambda,V)$.
One can separate the ultra-violet (UV) 
$2\int d \lambda \lambda f(\lambda, V)$ from its thermal
contributions $ \int d \lambda \lambda N(\lambda/a) f(\lambda,V)$.
Let us look at each terms more closely.

The UV term of $T^A_3$ is 
\begin{eqnarray}
T^A_{3UV} = && 
    \int^\Gamma d \lambda \lambda^2 \partial_v G_\infty(\lambda,V) 
       \partial_{v} G_\infty^*(\lambda,V)    \\
     &+& \frac{e_+^2}{8 \pi} e^{-2aV} \left( 
       e^{-i\beta_+^* V} \beta_+^* T^A_{3a} + e^{i \beta_+ V}
	\beta_+ T^{A*}_{3a} \right)+
	\frac{e_+^2}{16 \pi} e^{-2(\epsilon_++a) V} T^A_{3b}.
	\label{T^A_3},\nonumber
\end{eqnarray}
where $T^A_{3a}$ and $T^A_{3b}$ are given by
Eqs. (\ref{T3a:int}) and (\ref{T3b:int}) if one replace 
$\omega \rightarrow \lambda $, $v \rightarrow V$.
The first term of $T^A_3$
is canceled by the $\int d \lambda \lambda 
 \left\{\partial_vG_\infty(\lambda,V) \partial_{v} \xi^*_\lambda(V)
+\partial_v \xi _\lambda(V) \partial_{v}G_\infty^*(\lambda,V)\right\}$
term of $T^A_2$ \cite{massar}.
The dominant term for this UV contribution is 
\begin{eqnarray}\label{T_3^AUV}
 \frac{e_+^2}{4 \pi}&& \left(\frac{1-e_-/e_+}{m \omega_+}
     \right)^2 \ln \frac{\Gamma}{\omega_0}  
 \omega_0^2 \cos^2(\omega_+ V + \theta)
	  e^{-2(\epsilon_++a) V} \\
 &&=\frac{e_+^2}{16 \pi}\left(\frac{1-e_-/e_+}{m \omega_+} 
	\right)^2 \ln \frac{ \Gamma}{\omega_0} 
	\left[ 2 \omega_0^2 + \beta_+^2 (av)^{2i \frac{\omega_+}{
	a}} + \beta_+^{*2} (av)^{-2i \frac{\omega_+}{a}}
	\right] (av)^{-2(1+ \epsilon/a)} \nonumber
\end{eqnarray}
There are thermal contributions  in $T^A_3$ but we can
argue that it does not give comparable contribution to  the UV term.
The general form of the integral of the thermal part is 
\begin{eqnarray}
\int d \lambda  \frac{2\lambda^2}{e^{\lambda/a}-1} 
	  \partial _V G(\lambda, V) \partial_V G^*(\lambda, V).
\end{eqnarray} 
As one can easily see, there is no UV divergence because
of the thermal factor in the denominator. Moreover there is 
no IR contributions which comes from $\lambda \sim 0$.  Therefore
it do not give  terms depends on the cutoff $\Gamma$ which is the
main contribution of the $T^A_{3UV}$. 

In case of  $T^A_2$  the situation is much different to $T^A_3$ because
there are no UV contributions and the main contribution of it is only
for small $V$. So we must calculate it exactly in the region 
$\frac{1}{\Gamma} < V \ll 1/\omega_0,  2/a$. $T_2^A$ 
is sum of two terms  which are mutually complex conjugate.
One of these is 
\begin{eqnarray}
&&\int d \lambda \lambda  \coth (2\pi \lambda/a)  \partial_V G(\lambda, V) 
	\partial_V \xi_{\lambda}^*(V) e^{-2a V}    \\
&&=  \int d \lambda \lambda \coth (2\pi \lambda/a)  \partial_V G_\infty(\lambda, V) 
	\partial_V \xi_{\lambda}^*(V) e^{-2aV}
       + \frac{i e_+}{8 \pi}
	e^{-(\epsilon_+ + 2 a)v} T^A_{2a}, \nonumber
\end{eqnarray}
where
\begin{eqnarray}\label{T^A_{2a}}
T^A_{2a} &=&  -\beta_+ e^{i\omega_+V} \int d \lambda \lambda
	\coth{2 \pi \lambda/a }\chi_+^*(-\lambda) e^{i\lambda V}  \\
	&+& \beta_+^* e^{-i \omega_+V}
	\int d\lambda \lambda \coth{2 \pi \lambda /a} \chi_+(\lambda) 
		e^{i \lambda V} . \nonumber
\end{eqnarray}
Therefore we must calculate $T^A_{2a}- T^{A*}_{2a}$.  After change
of variable and using the fact $\lambda \coth{2 \pi \lambda/a}  $
is even function on $\lambda$, we get
\begin{eqnarray}
T^A_{2a} &-& T^{A*}_{2a} = -\beta_+ e^{-i \omega_+ V} \int_{-\infty}^{\infty}
	d \lambda \left[ \lambda \coth{2 \pi \lambda/a} \right] 
	\xi_+^*(-\lambda)e^{i \lambda V}    \\
	&-& C.C. \nonumber
\end{eqnarray}
Now we use 
\begin{eqnarray}
\coth \pi x = \frac{1}{\pi x} + \frac{2 x}{\pi}\sum_{k=1}^{\infty} \frac{1}{x^2+ k^2}
\end{eqnarray}
and do the residue integral along the upper half plane of the $\lambda$
plane, then we get
\begin{eqnarray}
T^A_{2a} &-& T^{A*}_{2a} = ia \beta_+ e^{-i \omega_+ V} \left[
	S(-2i \beta_+) - \frac{e_-}{2 e_+} \left\{ \left(1+ \frac{\omega_+}{\omega_-}
		\right) S(-2i \beta_-) + \left(1- \frac{\omega_+}{\omega_-}\right)
	 	S(2i \beta_-^*) \right\} \right]   \nonumber \\
      &-& C.C
\end{eqnarray}
where 
\begin{eqnarray}
S(\beta)= \sum_{k=1}^{\infty} \frac{ak e^{-akV/2}}{ak + \beta}.
\end{eqnarray}
If we restrict $V$ to $V \ll 2/a, 1/\omega_0$,  we get
\begin{eqnarray}
S(\beta)  \cong \frac{2}{a V} + \frac{\beta}{a} \ln\left(\frac{V}{\omega_0}\right).
\end{eqnarray}
The second term is much smaller than the first. Therefore we can write
\begin{eqnarray}
T^A_2  &=& - \frac{e_+}{4 \pi} \left(1- \frac{e_-}{e_+} \right)
		e^{-(\epsilon_+ + 2a)V } \frac{\omega_0}{V} \cos(\omega_+ V+ \theta) \\
   &=& - \frac{e_+}{4 \pi m \omega_+} \left( 1-
     \frac{e_-}{e_+}    \right) \frac{ a}{
	\ln av} \left[ \beta_+ (av)^{i \frac{
	\omega_+}{a}} + \beta_+^* ( av)^{-i \frac{\omega_+}{a}}
	\right] (av)^{-(2+ \epsilon_+/a)}. \nonumber
\end{eqnarray}

\end{document}